\documentclass[twocolumn,showpacs,preprintnumbers,amsmath,amssymb]{revtex4}

\usepackage{amsmath}
\usepackage{graphicx}
\usepackage{epsf}
\usepackage{psfrag}
\usepackage{epsfig}
\usepackage{graphics}

\begin{document}

\hyphenation{re-nor-ma-li-za-tion}
\hyphenation{diffe-ren-ce}
\hyphenation{pro-ba-bi-li-ty}

\newcommand{\be}{\begin{equation}}
\newcommand{\ee}{\end{equation}}
\newcommand{\bq}{\begin{eqnarray}}
\newcommand{\eq}{\end{eqnarray}}
\newcommand{\dt}{\frac{d^3k}{(2 \pi)^3}}
\newcommand{\dtp}{\frac{d^3p}{(2 \pi)^3}}
\newcommand{\kbruto}{\hbox{$k \!\!\!{\slash}$}}
\newcommand{\elemintcut}{\int_{\Lambda} \frac{d^4 k}{(2\pi)^4}}
\newcommand{\elemint}{\int \frac{d^4 k}{(2\pi)^4}}
\newcommand{\elemintl}{\int \frac{d^4 l}{(2\pi)^4}}
\newcommand{\derbruto}{\hbox{$\partial \!\!\!{\slash}$}}
\newcommand{\pbruto}{\hbox{$p \!\!\!{\slash}$}}
\newcommand{\qbruto}{\hbox{$q \!\!\!{\slash}$}}
\newcommand{\lbruto}{\hbox{$l \!\!\!{\slash}$}}
\newcommand{\abruto}{\hbox{$a \!\!\!{\slash}$}}
\newcommand{\quiral}{e^{\imath \alpha \gamma_5}}
\newcommand{\quiralconj}{e^{-\imath \alpha \gamma_5}}
\newcommand{\duploquiral}{e^{2\imath \alpha \gamma_5}}
\newcommand{\duplok}{k_\mu k_\nu}
\newcommand{\duplop}{p_\mu p_\nu}
\newcommand{\pmu}{p_\mu}
\newcommand{\pnu}{p_\nu}
\newcommand{\kmu}{k_\mu}
\newcommand{\knu}{k_\nu}
\newcommand{\lmu}{l_\mu}
\newcommand{\lnu}{l_\nu}
\newcommand{\lnk}{\ln\left(-\frac{k^2}{\lambda^2}\right)}
\newcommand{\denk}{k^2 (k - p)^2}
\newcommand{\denl}{l^2 (k - l)^2}
\newcommand{\denkepsilon}{(k^2 - m^2)^{2 - \epsilon} [(k - p)^2 - m^2]}
\newcommand{\pdois}{p^2(1 - 2x)}
\newcommand{\Hdois}{p^2x(1 - x) - m^2}
\newcommand{\mdois}{m^2}
\newcommand{\fracp}{\frac{\imath}{\pbruto_n + \kbruto - m}}
\newcommand{\Ilog}{I_{log}(\lambda^2)}
\newcommand{\Ilogdois}{I^{(2)}_{log}(\lambda^2)}
\newcommand{\lnp}{\ln\left(-\frac{p^2}{\lambda^2}\right)}
\newcommand{\lndois}{\ln^2\left(\frac{p^2}{\lambda^2}\right)}
\newcommand{\lnpm}{\ln\left(-\frac{p^2}{m^2}\right)}
\newcommand{\eps}{\epsilon}

\title{{\bf{Implicit regularization beyond one loop order: gauge field theories}}}

\date{\today}

\author{E. W. Dias$^{(a)}$} \email []{edsondias@fisica.ufmg.br}
\author{A. P. Ba\^eta Scarpelli$^{(b)}$} \email[]{scarp@fisica.ufmg.br}
\author{L. C. T.  Brito}\email[]{lctbrito@fisica.ufmg.br}
\author{Marcos Sampaio$^{(a)}$}\email[]{msampaio@fisica.ufmg.br}
\author{M. C. Nemes$^{(a)}$}\email[]{carolina@fisica.ufmg.br}

\affiliation{(a) Federal University of Minas Gerais - Physics
Department - ICEx \\ P.O. BOX 702, 30.161-970, Belo Horizonte MG -
Brazil}
\affiliation{(b) Centro Federal de Educa\c{c}\~ao Tecnol\'ogica - MG \\
Avenida Amazonas, 7675 - 30510-000 - Nova Gameleira - Belo Horizonte
-MG - Brazil}

\begin{abstract}
\noindent
We extend a constrained version of Implicit Regularization (CIR) beyond one loop order for gauge field theories.
In this framework, the ultraviolet content of the model is displayed in terms of momentum loop
integrals order by order in perturbation theory for any Feynman diagram, while the
Ward-Slavnov-Taylor identities are controlled by finite surface terms.  To illustrate, we apply CIR to massless abelian Gauge Field Theories (scalar and spinorial QED) to two loop order and calculate the two-loop
beta-function of the spinorial QED.
\end{abstract}


\maketitle

\section{Introduction}

Alternatives to dimensional regularization (DR) need to be constructed for quantum field theoretical
models which are well-defined only in their physical dimension. For instance, in supersymmetric theories
the analytical continuation of the space-time dimension from $4$ to $d$ for the vector fields leads to a
mismatch between the fermionic and bosonic degrees of freedom. This gives rise to a breaking of the supersymmetric relations.
A fully consistent and symmetry preserving  regularization and renormalization framework, valid to arbitrary loop order, has
not yet been constructed. Instead, a more pragmatical approach is employed by using Dimensional Reduction (DRed) \cite{DRED}.
Although DRed is mathematically inconsistent and cannot be extended to arbitrary loop order, it can be successfully applied if
it is judiciously used in order to preserve the supersymmetric Slavnov-Taylor identities of the model. In principle, an invariant
regularizaton scheme could be precluded by adopting an strategy in which  symmetry breakings are compensated by symmetry restoring
counterterms, once genuine (physical) quantum symmetry breakings have been discarded in the underlying renormalizable model.
In practice, a more efficient approach would be provided by a manifestly supersymmetric and gauge invariant regularization.
From a more phenomenological perspective it is expected that the Minimal Supersymmetric Standard Model (MSSM) can be probed at
LHC with an accuracy at the percent level in measurements of  electroweak precision observables \cite{EWPO}. This, on
the theory side, is realized through virtual effects of supersymmetric particles.  The theoretical evaluation of such
observables must be performed at least to two loop order both in the SM and MSSM so that the evidence of new physics becomes
unraveled. Clearly, an invariant regularization and renormalization procedure  is important, both to eliminate
inconsistencies and to reduce the number of adjustable parameters in the calculation, particularly when a supersymmetry breaking
mechanism is taken into account in more phenomenological physical models.

Implicit regularization (IR) technique \cite{IR1}-\cite{IR15} is a relatively new momentum space framework which
operates in the physical dimension of the theory. Under the assumption that some regulator is implicitly being
used  in a divergent amplitude, it is possible to algebraically separate the divergences in terms of momentum loop
integrals which are independent of external momenta, and may be directly absorbed in the definition of the renormalization
constants. In this framework, a minimal subtraction consists in defining the renormalization constants using mass
independent loop integrals. This is achieved by exchanging the mass dependence for  an  arbitrary non-vanishing scale by
means of a regularization independent identity. Such scale  parametrizes the freedom of separating the divergent part of
an amplitude and plays the role of renormalization group scale.  IR can be generalized to arbitrary loop
order \cite{IR2},\cite{IR9}. Assuming that the theory has been renormalized to $n-1$ loop order, it is always possible to
cast the ultraviolet behavior in $n$ loop order as a momentum loop integral. In other words IR is compatible with
the BPHZ forest formula which defines the set of subtractions to remove subdivergences. The derivatives of the renormalization
constants which are useful to calculate renormalization group functions can be displayed as loop integrals as well. Because the
loop basic divergent integrals are not evaluated, nor is the integrand of the amplitude modified, such procedure neatly separates
the regularization dependent part (loop momentum integrals) from the finite part (which can be evaluated by standard methods
\cite{SMIRNOV}) in a unitary preserving fashion. IR has been  shown to be a symmetry preserving framework. We have verified that
symmetry breaking terms stem from well defined finite differences between loop divergent integrals of the same superficial degree
of divergence. Those differences are arbitrary (regularization dependent) numbers which can be written as surface terms and appear
matter-of-factly in the process of defining a complete set of basic divergent integrals within IR. Yet such symmetry
breaking surface terms can be handled by either choosing finite counterterms to restore the symmetry of the renormalized
Green's functions,  or by adding local counterterms into the Lagrangian so that the Slavnov-Taylor identities are confirmed,
an invariant scheme is certainly more appealing. A constrained version of IR (CIR) in which those surface terms are
set to zero ab initio has been shown to automatically preserve abelian and non-abelian gauge symmetry  to one loop order
and supersymmetry to three loop order \cite{IR5},\cite{IR6},\cite{IR7},\cite{IR9},\cite{IR10},\cite{IR12}. It is noteworthy
that such surface terms vanish if they are computed in Dimensional Regularization which is notoriously a gauge invariant scheme
though it breaks supersymmetry due to the space-time dimensional continuation. CIR was mapped onto Constrained Differential
Renormalization \cite{DIFF}, a coordinate space perturbative renormalization program which automatically delivers
Green's functions as functions of a single mass (renormalization group)
scale that  fulfill the corresponding Ward-Slavnov-Taylor identities
of the underlying model. Although Differential Renormalization has been successfully applied to higher order
calculations, its constrained version has not been established yet \cite{SEIJAS}.

It is instructional to discuss the role of the surface terms that appear within IR in respecting
the symmetry of a model on the quantum level. We have verified that setting surface terms to zero automatically
ensures momentum routing invariance in a Feynman diagram \cite{IR5},\cite{IR6}. In other words, shifts in the
integration variables will be allowed. On the other hand it is well known that the diagrammatic proof of the
Ward-Takahashi identity for correlation functions in QED involves a shift in the integration variable \cite{PESKIN}
which is always legitimated in Dimensional Regularization. Moreover when anomalies (quantum symmetry breakings) are
present they manifest themselves as a breaking of momentum routing invariance in perturbation theory \cite{JACKIW}.
IR correctly displays the anomaly among the Ward identities in a democratic way just as a sound regularization framework
should by letting physics to fix the identity to be violated \cite{IR5},\cite{IR6},\cite{IR11}. A good example of this
feature is the AVV triangle anomaly: the conservation of the vector or axial-vector current depends on the context it
arises as the answer is not intrinsic to the AVV triangle graph. In such cases, as well as when radiative corrections appear
to be finite and undetermined \cite{JACKIW2}, surface terms are left as free parameters to be fixed on physical grounds.

The idea of associating surface terms and variable of integration ambiguities to the preservation of conventional
and super Ward-Takahashi-Slavnov Taylor identities is not new. In a program called Pre-Regularization \cite{MCKEON},
it was verified in some examples that a particular choice of the momentum shift parameters in conjunction with
dimensional regularization could be used to preserve gauge and supersymmetry. IR goes somewhat in
the opposite direction. The vanishing of surface terms which also leads to momentum routing invariance turns up
to preserve gauge and supersymmetry systematically.

In a previous contribution, we have shown that IR can be applied to arbitrary loop order by making use
of the BPHZ forest formula to remove subdivergences \cite{IR2}. Basic divergent integrals, as well as new
surface terms are consistently defined to $n^{th}$ loop order. Given the discussion above, it is natural to
ask whether CIR respects gauge symmetry to arbitrary loop order. Bearing in mind the Ward-Takahashi identity
which states that if ${\cal{M}}=\epsilon^\mu M_\mu$ is an amplitude contributing to the S-matrix
($\epsilon^\mu$ being the polarization vector of an external gauge boson), then $k^\mu M_\mu = 0$, for this external
gauge boson with momentum $k_{\mu}$, the answer to this question is positive. Such statement stems from the fact that
gauge invariance and unitarity require that the non-transverse degree of freedom of any massless gauge boson should not
contribute to physically measurable quantities. The general diagrammatic proof of the Ward-Takahashi identity assumes
the possibility of a shift in the integration variable. Within CIR this is achieved automatically by setting surface terms
to zero from the start. In other words, an invariant regularization and renormalization scheme emerges by systematically
setting the surface terms to zero. To illustrate, we use scalar and spinorial QED as testing grounds.

The paper is organized as follows: In section II,
we present the procedure of CIR to n-loop order, discussing the strategy to subtract the divergent contributions. We also
illustrate the procedure with
a three-loop order calculation in the context of spinorial QED. Section III is reserved for a discussion on
the preservation of gauge symmetry and unitariry by Constrained Implicit Regularization.  In Section IV, we apply
CIR to the calculus of the vacuum polarization tensor in scalar QED. Section V is dedicated to spinorial
QED, also to two-loop order: the calculation of the vacuum polarization tensor, of the electron self-energy and of
the electron-photon-electron vertex correction are performed. The Ward-Takahashi identities are verified and the
beta-function is calculated. Section VI is dedicated to the conclusions. Some technical points are left to the appendix.

\section{n-loop Constrained Implicit Regularization}

In this section we state the basic steps of CIR to $n$-loop order.
The goal is to identify the typical divergence of the $n^{th}$ order and the finite part of an amplitude once
the model has been renormalized to $(n-1)^{th}$ order.

\begin{enumerate}

\item In order to give mathematical rigor to any algebraic
manipulation performed in the amplitude, we implicitly assume that a
regularization has been applied. It  can be maintained implicit, the only requirement
being that  the integrand of the amplitude nor the dimension of the
space-time is modified. A good one is the simple ultraviolet momentum cutoff. As we will
see, the possible symmetry violations are restored by construction.
After performing symmetry group operations,  we cast the momentum-space
amplitude as a  combination of {\it{basic integrals}}.

\item Once the basic integrals are obtained, the divergent part is  separated  as
{{\it loop divergent integrals}}, which are obtained
by applying recursively the algebraic identity,
\bq
&&\frac {1}{(p-k)^2-m^2}=\frac{1}{(k^2-m^2)} \nonumber \\
&&-\frac{p^2-2p \cdot k}{(k^2-m^2) \left[(p-k)^2-m^2\right]},
\label{ident}
\eq
in all propagators that depends on external momenta,
until the divergent part is freed from the external momentum
dependence in the denominator.
This will assure local counterterms.

\item Loop divergent integrals for which the internal momenta carry Lorentz indices are expressed as
functions of {\it {basic divergent integrals}} and surface terms (the n-volume integral of a total n-divergence).
Such surface terms are regularization dependent (e.g. they vanish in
dimensional regularization) and are connected to momentum routing invariance: local momentum routing dependent terms
show up in the evaluation of an amplitude multiplied by (dimensionless) surface terms, which may break gauge symmetry.
CRI assumes that such surface terms are canceled  by local symmetry restoring
counterterms. In practice, this is automatically realized by setting
them to zero from the start.  Care must be exercised when quantum symmetry breaking  occurs. In this case the surface
terms should be considered as finite arbitrary parameters to be fixed on
physical grounds. That is because anomalies are somewhat connected to momentum
routing dependence in Feynman diagram calculations \cite{JACKIW}.

\item The basic divergent integrals which encode the ultraviolet
behavior of the amplitude need not be evaluated. We adopt the following
notation for  basic divergent integrals of $n^{th}$ loop order:
\be I^{(n)}_{log}(m^2)=  \int_k^\Lambda
 \frac{1}{(k^2-m^2)^2}Z_0^{(n-1)}(k^2,m^2,\lambda^2)
\label{ilog}
\ee
\be
I^{(n)}_{quad}(m^2)=
\int_k^\Lambda \frac{1}{(k^2-m^2)}Z_0^{(n-1)}(k^2,m^2,\lambda^2)
\label{iquad}
\ee
for logarithmically and  quadratically basic
divergent integrals, etc. . Hereafter the superscript $\Lambda$ indicates that integral is regularized and
$\int_k \equiv \int d^4k/(2 \pi)^4$. The functions $Z_0^{(n-1)}$ are typical
finite terms of the $(n-1)^{th}$-loop order. $\lambda$ is a non-vanishing arbitrary parameter with mass dimension
which appears when we define a mass independent scheme as we explain below. At one-loop order
the basic divergent integrals in four dimensional space-time  reads $I_{log}(m^2) =
\int_k (k^2-m^2)^{-2}$, $I_{quad}(m^2) = \int_k (k^2-m^2)^{-1}$, etc.

\item The subtraction of the subdivergences is performed in the light of the BPHZ
Forest Formula \cite{BPHZ}. We adopt the version of  the forest
formula which is analogous to the ordinary counterterm method. The
subtraction operators are translated into local counterterms which
substitutes the subgraph.  This is implemented via a recursion equation
which involves disjoint renormalization parts only, described in
textbooks \cite{MUTA}.
The basic divergent integrals can be subtracted as they stand in the definition
of renormalization  constants. A minimal, mass independent scheme is defined by
substituting $m^2$ with $\lambda^2 \ne 0$  using a kind of scale relation which is regularization
independent. This is achieved by means of the identity,
\bq
&& \frac {1}{(k^2-m^2)^w}=\frac {1}{(k^2-\lambda^2)^w}
-(\lambda^2-m^2) \nonumber \\
&&\times \sum_{i=1}^w \frac
{1}{(k^2-m^2)^i(k^2-\lambda^2)^{w-i+1}}.
\eq
For a basic logarithmic integral in one loop order, such scale relation is
\be
\label{scale}
I_{log}(m^2)=I_{log}(\lambda^2)+b
\ln{\left( \frac{\lambda^2}{m^2}\right)},
\ee
where $b=\frac{i}{(4\pi)^2}$.  The object to be subtracted is
$I^{(n)}_{log}(\lambda^2)$, $\lambda$ playing the role of
 renormalization group scale. For infrared safe
massless models a systematic cancelation of $ln(m^2)$ coming from
(\ref{scale}) and from ultraviolet finite part will render the
amplitude well defined as $m^2 \rightarrow 0$.

\item The remaining ultraviolet finite integrals are evaluated as
usual using Feynman parameters  or other methods in momentum space \cite{SMIRNOV}.

\end{enumerate}

\begin{figure}
\resizebox{1\hsize}{!}{\includegraphics*{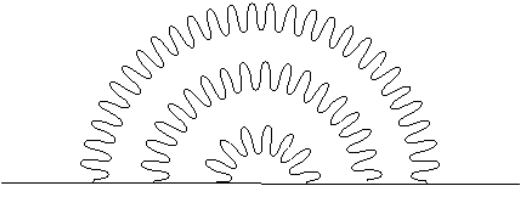}}
  \caption{a contribution for the three loop massless QED self-energy}
  \label{fig:1}
\end{figure}

We start off with some examples. In order to illustrate the role of surface terms in the preservation
of symmetries, we recall the one-loop massless QED vacuum polarization tensor, as
calculated in \cite{IR3}, with arbitrary momentum in the internal lines, $k_1$ and
$k_2$. It is written as:
\bq
\Pi_{\mu \nu}&=&\Pi(p^2)(p_\mu p_\nu-p^2 g_{\mu \nu}) \nonumber \\
&+& 4\left( \alpha_1 g_{\mu\nu}-\frac{1}{2}(k_1^2+k_2^2)\alpha_2 g_{\mu
\nu} \right. \nonumber \\
&+& \left.
\frac{1}{3}(k_{1}^{\alpha}k_{1}^{\beta}+k_{2}^{\alpha}k_{2}^{\beta}
+k_{1}^{\alpha}k_{2}^{\beta})
\alpha_3 g_{\{\mu \nu}g_{\alpha \beta\}}  \right.\nonumber \\ &-&
\left.
(k_1+k_2)^{\alpha}(k_1+k_2)_{\mu}\alpha_2 g_{\nu
\alpha} \right. \nonumber \\
&-& \left.
\frac{1}{2}(k_1^{\alpha}k_1^{\beta}+k_2^{\alpha}k_2^{\beta})g_{\mu \nu}
\alpha_2 g_{\alpha \beta}  \right).
\label{QED}
\eq
In the equation above, $p=k_1-k_2$ is the external momentum and
$$
\Pi(p^2)= \frac{4}{3} \Big[ I_{log}(\lambda^2) - b
\ln\Big(- \frac{p^2}{\lambda^2}\Big)+ \frac{5}{3} b\Big],
$$
with $b=i/(4 \pi)^2$, includes the basic divergent integral (in which we have used \ref{scale}). We have chosen the massless
limit just for the sake of simplicity. Now, the momentum routing dependent
terms, which lead to violation of gauge symmetry, are proportional to the $\alpha_i$'s, namely
\be
\alpha_1 g_{\mu \nu} \equiv \int^{\Lambda}_k
\frac{g_{\mu\nu}}{k^2-m^2}-
2\int^{\Lambda}_k
\frac{k_{\mu}k_{\nu}}{(k^2-m^2)^2},
\ee
\be
\alpha_2 g_{\mu \nu} \equiv \int^{\Lambda}_k
\frac{g_{\mu\nu}}{(k^2-m^2)^2}-
4\int^{\Lambda}_k
\frac{k_{\mu}k_{\nu}}{(k^2-m^2)^3}
\label{CR1}
\ee
and
\bq
\alpha_3 g_{\{\mu \nu}g_{\alpha \beta\}}  & \equiv &
g_{\{\mu \nu}g_{\alpha \beta \}}
\int^{\Lambda}_k
\frac{1}{(k^2-m^2)^2} \nonumber \\
&-&24\int^{\Lambda}_k
\frac{k_{\mu}k_{\nu}k_{\alpha}k_{\beta}}{(k^2-m^2)^4}.
\label{CR2}
\eq
These parameters are surface terms. It can be easily shown that
\be
\alpha_2 g_{\mu \nu}= \int_k ^\Lambda \frac{\partial}{\partial k^\mu}
\left( \frac{k_ \nu}{(k^2-m^2)^2} \right),
\ee
\be
\alpha_1 g_{\mu \nu}= \int_k ^\Lambda \frac{\partial}{\partial k^\mu}
\left( \frac{k_ \nu}{(k^2-m^2)} \right)
\ee
and
\be
\int_k^\Lambda \frac{\partial}{\partial k^\beta}
\left[ \frac{4k_\mu k_\nu k_\alpha}{(k^2-m^2)^3} \right]
=g_{\{\mu \nu}g_{\alpha \beta\}}(\alpha_3-\alpha_2).
\ee
As a second example illustrating the steps of CIR at $n$-loop order, let us now evaluate
the nested three-loop contribution to the fermion propagator in
massless QED as depicted in fig. \ref{fig:1}.

We begin with the  one-loop contribution, that reads
 \be
 i\Sigma(p)= -e^2 \int_k^\Lambda \frac{\gamma^\rho \kbruto
 \gamma_\rho}{k^2(p-k)^2}
\ee
and which may as well be written like
\be
i\Sigma(p)= 2 e^2 \gamma^\alpha I_\alpha,
\ee
where $I_{\alpha}$ is a linearly divergent integral
\be
I_\alpha=\int_k^\Lambda \frac{k_\alpha}{k^2(p-k)^2}.
\ee
In order  to calculate this basic divergent integral, we apply the identity (\ref{ident})
twice to obtain
\bq
&&I_\alpha = \int_k^\Lambda \frac {k_\alpha}{(k^2-m^2)}\left\{ \frac
 {1}{(k^2-m^2)} \right. \nonumber \\
&&\left. -\frac{p^2-2p\cdot k}{(k^2-m^2)^2}
+\frac{(p^2-2p\cdot k)^2}{(k^2-m^2)^2[(p-k)^2-m^2]}\right\}, \nonumber
 \\
&&
\eq
in which the limit $m^2 \to 0$ will be taken in the end of the
calculation. Thus
\bq
&&I_\alpha= 2p^\mu \int_k^\Lambda \frac{k_\mu k_\alpha}{(k^2-m^2)^3}
 \nonumber \\
&&+ \int_k \frac{k_\alpha (p^2-2p\cdot k)^2}{(k^2-m^2)^3[(p-k)^2-m^2]}
\nonumber \\
&&=\frac {p^\mu}{2} \left( - \alpha_2 g_{\mu \alpha}   \right. \nonumber \\
&&\left. + g_{\mu \alpha}
\int_k^\Lambda
\frac {1}{(k^2-m^2)^2}\right)+ \tilde I_\alpha (p^2,m^2),
\eq
$\tilde I_\alpha$ being a finite integral. In the limit $m^2 \to 0$ we obtain, taking (\ref{scale}) into account
\bq
&&I_\alpha= \frac {p_\alpha}{2} \left(I_{log}(\lambda^2)-b
\ln{\left(-\frac{p^2}{\lambda^2}\right)}+2b\right) \nonumber \\
&& -\frac{p^\mu}{2} \alpha_2 g_{\mu \alpha},
\eq
where $\alpha_2$ is the surface term defined in (\ref{CR1}) that will be set to zero. Thus
\be
i\Sigma(p)=  e^2 \pbruto \left\{ I_{log}(\lambda^2)-b
\ln{\left(-\frac{p^2}{\lambda^2}\right)}+2b\right\}
\ee
Now consider  the nested two-loop self-energy. According to the BPHZ forest
formula the subtraction of the one loop subdivergence amounts to replace the inner diagram with its finite part, namely
\bq
&&i\Sigma^{(2)}_1(p)= -i b e^4 \int_k^\Lambda \frac{\gamma^\rho
 \kbruto \gamma_\rho}{k^2(p-k)^2}
\left\{\ln{\left(-\frac{k^2}{\lambda^2}\right)}+2\right\} \nonumber \\
&&=2ie^4b \gamma^\alpha \left[ 2I_\alpha -I_\alpha^{(2)}\right].
\eq
In the equation above
\be
I^{(2)}_\alpha=\int_k^\Lambda \frac{k_\alpha}{k^2(p-k)^2}
 \ln{\left(-\frac{k^2}{\lambda^2}\right)}.
\ee
Again we apply identity  (\ref{ident}) twice, to obtain
\bq
&&I^{(2)}_\alpha=\int_k^\Lambda  \frac
 {k_\alpha}{(k^2-m^2)}\ln{\left(-\frac{k^2-m^2}{\lambda^2}\right)}
\left\{ \frac {1}{(k^2-m^2)} \right. \nonumber \\
&&\left. -\frac{p^2-2p\cdot k}{(k^2-m^2)^2}
+\frac{(p^2-2p\cdot k)^2}{(k^2-m^2)^2[(p-k)^2-m^2]}\right\}. \nonumber
 \\
&&
\eq
Following the same steps as before, we have:
\bq
&&I^{(2)}_\alpha= 2p^\mu \int_k^\Lambda \frac{k_\mu
 k_\alpha}{(k^2-m^2)^3}\ln{\left(-\frac{k^2-m^2}{\lambda^2}\right)} \nonumber \\
&&+ \int_k \frac{k_\alpha (p^2-2p\cdot
 k)^2}{(k^2-m^2)^3[(p-k)^2-m^2]}\ln{\left(-\frac{k^2-m^2}{\lambda^2}\right)}
\nonumber \\
&&=\frac {p^\mu}{2} \left( -\int_k^\Lambda \frac{\partial}{\partial
k^\mu}
\left(
 \frac{k_\alpha}{(k^2-m^2)^2}\ln{\left(-\frac{k^2-m^2}{\lambda^2}\right)}\right)    \right. \nonumber \\
&&\left. + g_{\mu \alpha}
\int_k^\Lambda
\frac
 {1}{(k^2-m^2)^2}\ln{\left(-\frac{k^2-m^2}{\lambda^2}\right)}\right. \nonumber \\
&&\left. +\frac {g_{\mu \alpha}}{2} \int_k^\Lambda \frac
 {1}{(k^2-m^2)^2}\right)
+ \tilde I^{(2)}_\alpha,
\eq
$\tilde I^{(2)}_\alpha$ being a finite integral. A cancelation of divergent
logarithms between the ultraviolet finite and divergent parts as $m^2 \to 0$ takes place to render the amplitude well defined, namely
\bq
&&I^{(2)}_\alpha=\frac {p_\alpha}{2}\left\{I^{(2)}_{log}(\lambda^2)+
 \frac 12 I_{log}(\lambda^2)\right. \nonumber \\
&& \left. -\frac b2\left[  \ln^2{\left(-\frac{p^2}{\lambda^2}\right)}
- \ln{\left(-\frac{p^2}{\lambda^2}\right)} -3 \right] \right\},
\eq
in which we have adopted the following notation for the basic
logarithmically divergent two-loop integral
for massless theories,
\be
I^{(2)}_{log}(m^2) \equiv \int_k^\Lambda
 \frac{1}{(k^2-m^2)^2}\ln{\left(-\frac{k^2-m^2}{\lambda^2}\right)},
\ee
and we have used  the two-loop scale relation
\bq
&&I^{(2)}_{log}(m^2)=I^{(2)}_{log}(\lambda^2) \nonumber \\
&&-b\left\{ \frac 12 \ln^2{\left(\frac{m^2}{\lambda^2}\right)}
+ \ln{\left(\frac{m^2}{\lambda^2}\right)}\right\}.
\eq
Altogether we get, for the two loop nested electron self energy
\bq
&&i\Sigma^{(2)}_1(p)=  ie^4b \pbruto \left\{ -
 I^{(2)}_{log}(\lambda^2) + \frac 32  I_{log}(\lambda^2) \right. \nonumber \\
&&\left. + \frac b2  \ln^2{\left(-\frac{p^2}{\lambda^2}\right)}
- \frac 52 b \ln{\left(-\frac{p^2}{\lambda^2}\right)} +\frac 52 b
 \right\}.
\label{2loop}
\eq

Finally, we calculate the three-loop nested self-energy. Again, according to the forest formula,
to compute the divergence of the order, the subtraction of the subdivergences amounts to plugging
the finite part of the two-loop subgraph into the total three-loop amplitude, i.e.
\be
i \Sigma_1^{(3)}(p)=-ie^2\int_k^\Lambda
\frac{\gamma^\rho \kbruto\left(i\tilde \Sigma_1^{(2)}(\kbruto)\right)\kbruto \gamma_\rho}{k^4(p-k)^2},
\ee
where the tilde stands for the finite part of (\ref{2loop}). Therefore we are able to write
\be
i \Sigma_1^{(3)}(p)= -e^6b^2 \gamma^\alpha \left(  I_\alpha^{(3)} -  5 I_\alpha^{(2)}
+ 5 I_\alpha \right),
\ee
with
\be
I^{(3)}_\alpha=\int_k^\Lambda \frac{k_\alpha}{k^2(p-k)^2} \ln^2{\left(-\frac{k^2}{\lambda^2}\right)}.
\label{i3}
\ee
We proceed as before. In the integral above, we isolate the divergence
with the help of identity (\ref{ident}), which is applied twice to get
\bq
&&I^{(3)}_\alpha= 2p^\mu \int_k^\Lambda \frac{k_\mu k_\alpha}{(k^2-m^2)^3}\ln^2{\left(-\frac{k^2-m^2}{\lambda^2}\right)} \nonumber \\
&&+ \int_k \frac{k_\alpha (p^2-2p\cdot k)^2}{(k^2-m^2)^3[(p-k)^2-m^2]}\ln^2{\left(-\frac{k^2-m^2}{\lambda^2}\right)}
\nonumber \\
&&=\frac {p^\mu}{2} \left\{ -\int_k^\Lambda \frac{\partial}{\partial
k^\mu}
\left( \frac{k_\alpha}{(k^2-m^2)^2}\ln^2{\left(-\frac{k^2-m^2}{\lambda^2}\right)}\right)    \right. \nonumber \\
&&\left. + g_{\mu \alpha} \left(I_{log}^{(3)}(m^2) + I_{log}^{(2)}(m^2)
+ \frac 12 I_{log}(m^2)\right) \right\}  \nonumber \\
&&+ \tilde I^{(3)}_\alpha,
\eq
where
\be
I^{(3)}_{log}(m^2) \equiv \int_k^\Lambda \frac{1}{(k^2-m^2)^2}\ln^2{\left(-\frac{k^2-m^2}{\lambda^2}\right)}.
\ee
The complete result of (\ref{i3}) is achieved when the surface term is set to zero, the three-loop scale
relation,
\bq
&&I^{(3)}_{log}(m^2)=I^{(3)}_{log}(\lambda^2)
- b\left\{ \frac 13 \ln^3{\left(\frac{m^2}{\lambda^2}\right)} \right. \nonumber \\
&&\left. + \ln^2{\left(\frac{m^2}{\lambda^2}\right)} +2 \ln{\left(\frac{m^2}{\lambda^2}\right)} \right\},
\eq
is used and the finite part is calculated. We have
\bq
&&I^{(3)}_\alpha=  \frac {p_\alpha}{2} \left\{ I_{log}^{(3)}(\lambda^2) + I_{log}^{(2)}(\lambda^2)
+ \frac 12 I_{log}(\lambda^2) \right. \nonumber \\
&& \left. -\frac b3 \ln^3{\left(-\frac{p^2}{\lambda^2}\right)}
+\frac b2 \ln^2{\left(-\frac{p^2}{\lambda^2}\right)}  \right. \nonumber \\
&& \left.-\frac b2 \ln{\left(-\frac{p^2}{\lambda^2}\right)} +4b \right\}.
\eq
Finally, the three-loop contribution to the self-energy of fig. \ref{fig:1} is given by
\bq
&&i \Sigma_1^{(3)}(p)=-\frac{e^6 b^2 \pbruto}{2} \left \{
I_{log}^{(3)}(\lambda^2) -4 I_{log}^{(2)}(\lambda^2)  \right. \nonumber \\
&& \left. + 3 I_{log}(\lambda^2) -\frac b3 \ln^3{\left(-\frac{p^2}{\lambda^2}\right)}
+3b \ln^2{\left(-\frac{p^2}{\lambda^2}\right)} \right. \nonumber \\
&& \left. -8b \ln{\left(-\frac{p^2}{\lambda^2}\right)} + \frac {13}{2}b \right\}.
\eq
The procedure we described in this section will be used henceforth to compute
the two-loop scalar QED and spinorial QED. As it will be shown, the elimination of
surface terms plays essential role in satisfying the Ward identities.

\section{Discussion on Gauge Invariance and Unitarity}

In this section, we show that CIR is built so as to preserve gauge symmetry and, as a consequence, unitarity. We
are concerned here with QED, although this argument can be easily extended to the non-abelian case. As discussed
in the previous section, the prescription to be followed assumes that some regularization has been applied.
As we will see, this regularization does not necessarily have to preserve gauge symmetry, since gauge invariance
is restored by construction. So, we assume here that the regularization is a simple ultraviolet cutoff, which does not
modify neither the dimension of space-time nor the integrand.
The procedure we are adopting is the traditional diagrammatic proof of gauge invariance of
the text books (see, for example, ref. \cite{PESKIN}), with the addition of the regularization.

Let us consider the amplitude, ${\cal M}(k)=\epsilon_\mu (k) {\cal M}^\mu (k)$, for some QED process, involving
an external photon with momentum $k$. Then the Ward-Takahashi identity requires that
\be
k_\mu {\cal M}^\mu (k)=0.
\label{wi}
\ee
Since the external momenta are not necessarily on-shell, some of the contributions to the r.h.s of
(\ref{wi}) are non-null , but they do not contribute when the $S$-matrix element
is extracted. We can begin by any specific diagram that contributes to ${\cal M}$.
If the external photon $\gamma (k)$ is removed, we get a new diagram that contributes to a simpler amplitude,
${\cal M}_0$. If now we sum over  all possible insertions of this photon in ${\cal M}_0$, we must obtain
eq. (\ref{wi}).

Concerning the possibilities of insertion, we have two situations: the insertion can be made in an open fermion line or in
a closed fermion line. For the case of an open fermion line, there is nothing different from the procedure of
text books. These contributions are the ones which are non-null in the r.h.s of (\ref{wi}), but which do not
contribute when the $S$-matrix is considered. So, we dedicate ourselves to the case of insertions in a closed fermion line.

Let us consider this line has $n$ vertices. The fermion momenta for this line are $p_1$, $p_2$, $\cdots$,
$p_n$. If the insertion is done in a point with momentum $p_1$, all the fermion propagators after the insertion have
the momentum increased by $k$. We use the identity
\be
\kbruto= \left[(\pbruto_1+ \kbruto -m)-(\pbruto_1 -m)\right],
\ee
so that
\bq
&&\frac{i}{\pbruto_1 +\kbruto -m}(\kbruto)\frac{i}{\pbruto_1-m} \nonumber \\
&&  = i \left( \frac{i}{\pbruto_1-m}-\frac{i}{\pbruto_1 +\kbruto -m} \right).
\eq
The regularized integral over the loop momentum, $p_1$, takes the form
\bq
&&T^{\mu_1 \cdots \mu_n} \nonumber \\
&&=-i \int^\Lambda_{p_1} \mbox{tr}\left[ \frac{i}{\pbruto_n +\kbruto -m} \gamma^{\mu_n} \cdots
\frac{i}{\pbruto_2 +\kbruto -m}\gamma^{\mu_2}\right. \nonumber \\
&&\left.  \times  \left( \frac{i}{\pbruto_1-m}-\frac{i}{\pbruto_1 +\kbruto -m} \right)\gamma^{\mu_1} \right].
\eq
When the insertion is performed in all other points and the contributions are summed, there occurs many cancelations
and only two terms survive. We are left with
\bq
&&T^{\mu_1 \cdots \mu_n}  =-i \int^\Lambda_{p_1} \mbox{tr}\left[ \frac{i}{\pbruto_n  -m} \gamma^{\mu_n}
 \cdots  \frac{i}{\pbruto_1-m}\gamma^{\mu_1}\right. \nonumber \\
&& \left. - \frac{i}{\pbruto_n +\kbruto -m} \gamma^{\mu_n}
 \cdots  \frac{i}{\pbruto_1 +\kbruto -m}\gamma^{\mu_1} \right].
\eq
The two terms differs by a shift $p_1 \to p_1 +k$. If the integral is divergent, the cutoff regularization
does not allow a shift without compensating with a surface term. So, after shifting we get
\be
T^{\mu_1 \cdots \mu_n}=-i S^{\mu_1 \cdots \mu_n},
\ee
with $S^{\mu_1 \cdots \mu_n}$ being the surface term that causes violation of the Ward-Takahashi identity. The
prescription of Constrained Implicit Regularization is (whenever there are no anomalies) the restoration of
the symmetry by eliminating all the surface terms. CIR provides a simple algorithm to identify all the surface
terms. The procedure is explained by the rules 2 and 3 of the previous section: identity (\ref{ident}) is
recursively used in all propagators which depends on the external momenta until the divergent part
can be written in terms of loop integrals. The surface terms come from some of these loop integrals which have
Lorentz indices.

Concerning the unitarity, we call the reader's attention to the fact that, since the Ward-Takahashi identities
are respected, the amplitude under consideration, after the elimination of the surface terms and subdivergences, can be written as
\be
{\cal A}_{\mu_1 \cdots \mu_2}= L_{\mu_1 \cdots \mu_2} \left( {\cal A}^\Lambda +\bar {\cal A} \right).
\label{unit}
\ee
In the equation above, the tensor $L_{\mu_1 \cdots \mu_2}$ has the necessary structure so as to preserve
gauge invariance, ${\cal A}^\Lambda$ is the divergent part (cutoff dependent) and $\bar{\cal A}$ is the
finite part. The divergent piece contains only loop integrals, which will ask for local counterterms. From
(\ref{unit}) it is clear that the counterterms are also symmetric. So, unitarity is preserved.

\section{Explicit Calculations: Ward Identities in higher order calculations}

Ward identities are relations which need to be satisfied by the Green's functions in order to guarantee gauge invariance,
unitarity and renormalizability. From the perturbative standpoint, such relations must be satisfied
order by order and should not be spoiled by an invariant regularization.
In this section, we illustrate how CIR is a gauge invariant scheme by verifying some Ward identities.

\subsection{Vacuum polarization tensor in Scalar QED}
\begin{figure}
\resizebox{1\hsize}{!}{\includegraphics*{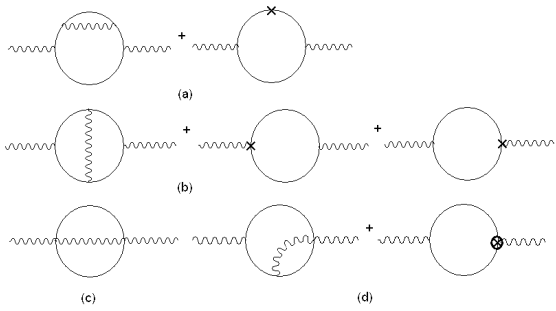}}
  \caption{two loop polarization vacuum tensor corrections in scalar QED}
  \label{fig:2}
\end{figure}

At two loop order, we have four non-vanishing contributions, in massless scalar QED,
for the vacuum polarization tensor, as shown in fig. \ref{fig:2}. Tadpoles diagrams do not contribute because quadratic
divergences can be shown to vanish in massless theories within IR \cite{IR2} just as in Dimensional Regularization.

In the diagram with nested subdivergences, fig. 2a, we can notice
that the subdiagram is the self-energy correction for the scalar
particle at one loop order. After applying the Feynman Rules for
scalar QED and subtracting the subdivergences with the forest
formula, we can write
\begin{eqnarray}
\Pi^a_{\mu \nu}(p^2) &=& (\imath e)^3(\frac{1}{\imath})^2\elemint \frac{(p +
               2k)_\mu(p + 2k)_\nu}{[(p + k)^2](k^2)^4}\nonumber\\
           & &
 \left[-2e^2k^2b\ln\left(\frac{k^2}{\lambda^2e^2}\right)\right]\mbox{.}
\end{eqnarray}
The amplitude can be displayed
in terms of typical basic integrals:
\begin{eqnarray}
\Pi^a_{\mu \nu}(p^2) &=& 2\imath e^4 b [p_\mu p_\nu I^{(2)}(p^2)} - 4p_\nu
 I_{\mu}^{(2)}{(p^2) + \nonumber\\
           & & 4I_{\mu\nu}^{(2)}(p^2) - 2\duplop I(p^2) - 8p_\nu
I_\mu(p^2) - \nonumber\\
           & & 8I_{\mu\nu}(p^2)]\mbox{,}
\end{eqnarray}
where
\be
I^{(i)},I^{(i)}_{\mu},I^{(i)}_{\mu\nu} =
 \int_k^\Lambda \frac{1, k_\mu, k_{\mu}k_{\nu}}{k^2(k - p)^2}\ln^{i-1}\left(-\frac{k^2}{\lambda^2}\right).
\ee
For $i = 1$ we simply write $I$, $I_{\mu}$, etc.

Next, we treat the contribution coming from fig. 2c. The subdivergences in this case are
proportional to basic quadratic divergences and, so, vanish for the massless case.
This amplitude reads
\bq
&&\Pi^c_{\mu \nu}(p^2) = -4\imath e^4g_{\mu\nu} \int_{k,l}^\Lambda \frac{1}{(p
 - k)^2l^2(l - k)^2}  \nonumber \\
 &&= -4\imath e^4g_{\mu\nu}\int_k^\Lambda \frac{1}{(p-k)^2} \nonumber \\
 && \times \left(I_{log}(\lambda^2)-b \ln\left(-\frac{k^2}{\lambda^2}\right)+2b \right),
\eq
where we have used the result of the one-loop basic integral $I$ in the internal momentum $l$.
In the equation above, taking into account that shifts are allowed in Implicit Regularization,
in our case only the term in the logarithm survives. This is because the others
are proportional to quadratic divergences. So, we are left with
\bq
&&\Pi^c_{\mu \nu}(p^2) = 4\imath b e^4g_{\mu\nu}\int_k^\Lambda \frac{1}{(p-k)^2}
\ln\left(-\frac{k^2}{\lambda^2}\right) \nonumber \\
&&= 4\imath b e^4g_{\mu\nu} J^{(2)}.
\eq
The next contribution is the one from fig. 2d:
\begin{eqnarray}
\Pi^d_{\mu \nu}(p^2)=
2\imath e^4\int^\Lambda_{k,l}\frac{(2k - p)_\mu (2k - l)_\nu}{k^2(k -
 p)^2l^2(k - l)^2}.
\end{eqnarray}
For this diagram, the elimination of the subdivergences is carried out by considering only the finite part
of the integral in the internal momentum $l$. So, using the results of the one loop integrals, we have
\bq
&&\Pi^d_{\mu \nu}(p^2)= 2\imath b e^4 \left(6 I_{\mu \nu} -\frac 32 p_\mu p_\nu I \right. \nonumber \\
&& \left. - 3 I^{(2)}_{\mu \nu} + \frac 32 p_\mu I^{(2)}_\nu \right).
\eq
In the expression above it was also taken into account that $I_\mu=p_\mu I/2$.

For the overlapped diagram of fig. 2b, we use the forest formula
to subtract the subdivergences. So, we have
\be
\Pi^b_{\mu\nu} = \Pi^{b1}_{\mu\nu} + 2\Pi^{bCT}_{\mu\nu}.
\ee
For the calculation of the first term, we have
\be
\Pi^{b1}_{\mu\nu}=-i e^4 \int_{k,l}^\Lambda \frac{(2k-p)_\mu(2k-p)_\nu M}
{k^2l^2(k-l)^2(p-l)^2(p-k)^2}
\ee
with
\bq
&&M=(k+l)\cdot(k+l-2p)  \nonumber \\
&&= 2\left\{ (l-p)^2+l^2-(k-l)^2-p^2 \right\}.
\eq
Thus, it is possible to write
\bq
&&\Pi^{b1}_{\mu\nu}= -2ie^4\left\{ -p^2(4 I_{\mu \nu}^{{\cal O}2}- p_\mu p_\nu I^{\cal O}) \right. \nonumber \\
&&\left. +\left[I_{log}(\lambda^2)+ 2b\right]\left[4I_{\mu \nu}-p_\mu p_\nu I\right] \right. \nonumber \\
&&\left. -2b\left[2I_{\mu \nu}^{(2)}-3p_\mu I_\nu^{(2)}+p_\mu p_\nu I^{(2)} \right]\right\},
\eq
where
\be
I^{\cal O},I_\mu^{\cal O},I_{\mu \nu}^{{\cal O}1}, I_{\mu \nu}^{{\cal O}2}=
\int_{k,l} \frac{1,k_\mu,k_\mu k_\nu, k_\mu l_\nu}
{k^2l^2(k-l)^2(p-l)^2(p-k)^2}.
\ee
Notice that the integrals above are symmetric in the exchange $k \leftrightarrow l$.
They will appear hereafter. Moreover
$I^{\cal O}$ and $I_\mu^{\cal O}=p_\mu I^{\cal O}/2$ are finite integrals.
For the counterterms, we have
\be
\Pi^{bCT}_{\mu\nu}=ie^4 I_{log}(\lambda^2)\left[4 I_{\mu \nu}- p_\mu p_\nu I\right],
\ee
and, so
\bq
&&\Pi^b_{\mu\nu} = -2ie^4\left\{ -p^2(4 I_{\mu \nu}^{{\cal O}2}- p_\mu p_\nu I^{\cal O}) \right. \nonumber \\
&&\left. -2b\left[2I_{\mu \nu}^{(2)}-3p_\mu I_\nu^{(2)}+p_\mu p_\nu I^{(2)} \right]\right\}.
\eq
We see that all the nonlocal divergent terms have been eliminated.
Adding all the contributions,
\begin{equation}
\Pi_{\mu\nu} = 2\Pi_{\mu\nu}^a + \Pi_{\mu\nu}^b + \Pi_{\mu\nu}^c +
4\Pi_{\mu\nu}^d,
\end{equation}
we are allowed to write
\begin{eqnarray}
&&\Pi_{\mu\nu} = \imath e^4 b\left\{- 16\pmu I^{(2)}_{\nu} + 8\duplop
I^{(2)} + 4 g_{\mu \nu} J^{(2)}\right. \nonumber\\
&& \left. +\frac{2p^2}{b}[4I^{{\cal O}2}_{\mu\nu} - \pmu\pnu I^{\cal O}]\right\}.
\end{eqnarray}
Writing explicitly each basic integral, using the results of the appendix,  we have
\begin{eqnarray}
&&\Pi_{\mu\nu} = (p^2g_{\mu\nu} - \duplop)\left\{4I_{log}(\lambda^2) - 4b \ln\left(-\frac{p^2}{\lambda^2}\right) \right. \nonumber\\
&&\left. + \frac{2}{b}\left[-\frac{1}{3}p^2I^{O} - p^2\frac{b^2\pi^2}{9}\right] + \frac{40}{3}\right\}.
\end{eqnarray}
So, we have obtained explicitly the transversal form of the vacuum polarization tensor
in two-loop scalar QED, as required by gauge invariance.
\subsection{QED Ward Identities to two-loop order}
\begin{figure}
\resizebox{1\hsize}{!}{\includegraphics*{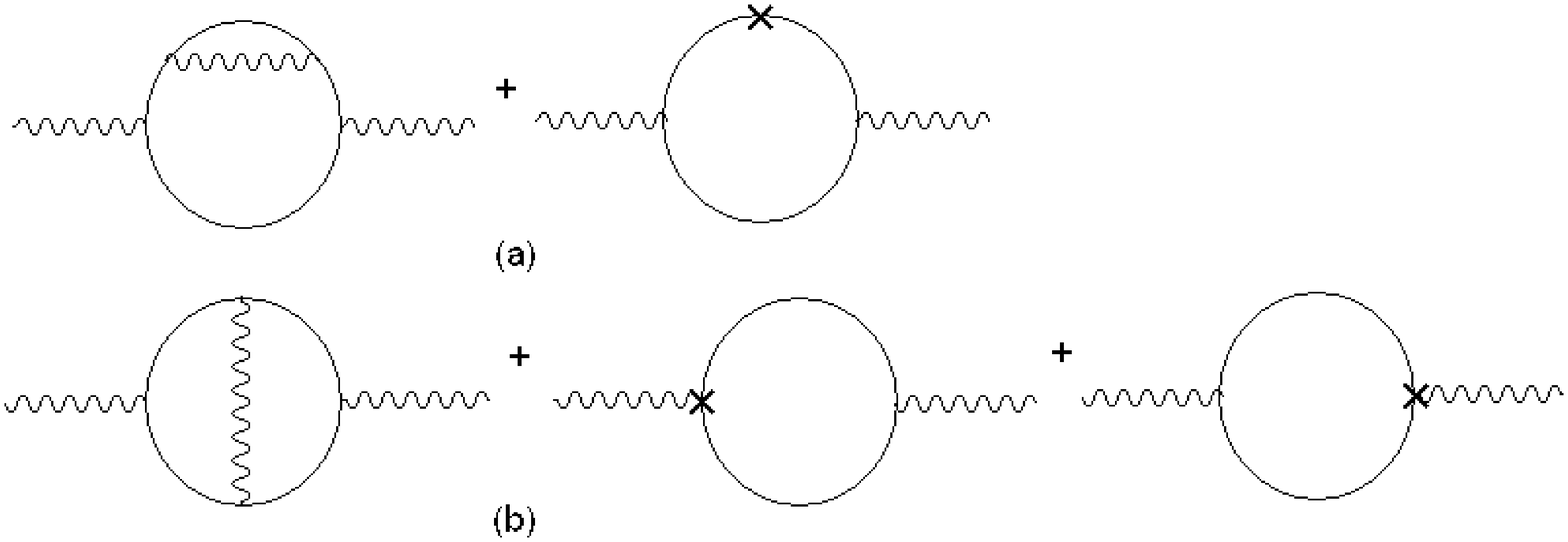}}
  \caption{Spinorial QED contributions for the polarization vacuum tensor at two loop order}
  \label{fig:3}
\end{figure}

We now apply CIR to spinorial QED to two loop order.
First, we shall obtain the vacuum polarization tensor, which has only two
contributions, as depicted in fig. \ref{fig:3}, where we also present the correspondent counterterms, that
eliminate subdivergences.

In the first diagram, the nested subdiagram  is the one-loop fermionic self energy correction. Taking only the finite
part of the subdiagram, we will correctly eliminate the subdivergences. We write
\bq
&&T^a_{\mu \nu} = \int_k^\Lambda \mbox{tr}\left\{(-ie \gamma_\mu) \frac {i}{\kbruto-\pbruto}
(-ie \gamma_\nu) \frac{i}{\kbruto} i\Sigma(\kbruto)\frac{i}{\kbruto}\right\} \nonumber \\
&&= ibe^4 \int_k^\Lambda \frac{\mbox{tr}(\gamma_\mu (\kbruto-\pbruto)\gamma_\nu \kbruto)}
{k^2(p-k)^2} \left(2- \ln\left(-\frac{k^2}{\lambda^2}\right)\right). \nonumber \\
&&
\eq
After some straightforward algebra, we have
\bq
&&T^a_{\mu \nu}=4ie^4b\left\{-2 I_{\mu \nu}^{(2)}+2p_\mu I_\nu^{(2)}+4I_{\mu \nu}  \right. \nonumber \\
&& \left. -2p_\mu p_\nu I -\frac{g_{\mu \nu}}{2} J^{(2)} +p^2 \frac{g_{\mu \nu}}{2}(2I-I^{(2)})\right\}
\eq

The second contribution corresponds to an overlapped diagram, for which we have to consider the two
equal counterterms represented in fig. 3b:
\be
T^b_{\mu \nu}=T^{b1}_{\mu \nu}+ 2T^{bCT}_{\mu \nu}.
\ee
For the first part, we have
\be
T^{b1}_{\mu \nu}=-ie^4\int_k^\Lambda \frac{\mbox{tr}
\{\gamma_\nu \lbruto \gamma^\rho \kbruto \gamma_\mu (\kbruto-\pbruto)\gamma_\rho(\lbruto-\pbruto)\}}
{k^2l^2(k-l)^2(p-k)^2(p-l)^2},
\ee
which, in terms of basic integrals, gives us
\bq
&&T^{b1}_{\mu \nu}=-ie^4\left\{ 2\left( I_{log}(\lambda^2)+2b\right)
\left(-I_{\mu \nu}+ p_\mu p_\nu I\right) \right. \nonumber \\
&&\left. +2b \left(I_{\mu \nu}^{(2)}- 2 p_\mu I_\nu^{(2)}\right) - \frac 12 p_\mu p_\nu (I)^2 \right. \nonumber \\
&&\left. + 2 p^2 I_{\mu \nu}^{{\cal O}1} - p^2p_\mu p_\nu I^{\cal O} \right. \nonumber \\
&&\left. +p^2 \frac {g_{\mu \nu}}{2} \left[ -4\left( I_{log}(\lambda^2)+2b\right)I+ p^2 I^{\cal O} \right. \right.
\nonumber \\
&& \left. \left. + (I)^2 +4b I^{(2)} \right] \right\}.
\eq
For the counterterms, we have
\be
2T^{bCT}_{\mu \nu}=8ie^4 I_{log}(\lambda^2)\left\{ 2I_{\mu \nu}-p_\mu p_\nu I+p^2 \frac {g_{\mu \nu}}{2} I\right\}.
\ee
Now, using the results of the appendix, we add these contributions in order to obtain the complete
spinorial QED vacuum polarization tensor:
\begin{eqnarray}
\label{eq: pimunutot}
T_{\mu\nu} &=& \frac{8}{3}\imath e^4 b(\duplop -
 g_{\mu\nu}p^2)\{\frac{3}{2}I^2_{log}(\lambda^2) -
                              3bI^{(2)}_{log}(\lambda^2) \nonumber\\
                          &+& \frac{31}{6}bI_{log}(\lambda^2) -
 \frac{3}{2}b\ln\left(-\frac{p^2}{\lambda^2}\right)\nonumber\\
                          &+& \frac{3}{2}b - p^2 I^{O}\}\mbox{.}
\end{eqnarray}

\begin{figure}
\resizebox{1\hsize}{!}{\includegraphics*{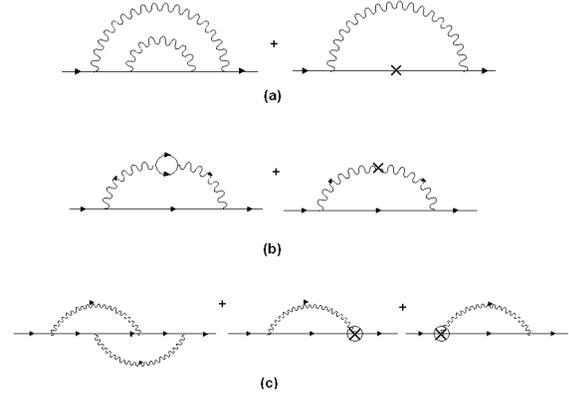}}
  \caption{contributions for the fermionic self energy at two loop order}
  \label{fig:4}
\end{figure}

\begin{figure*}
\resizebox{1\hsize}{!}{\includegraphics*{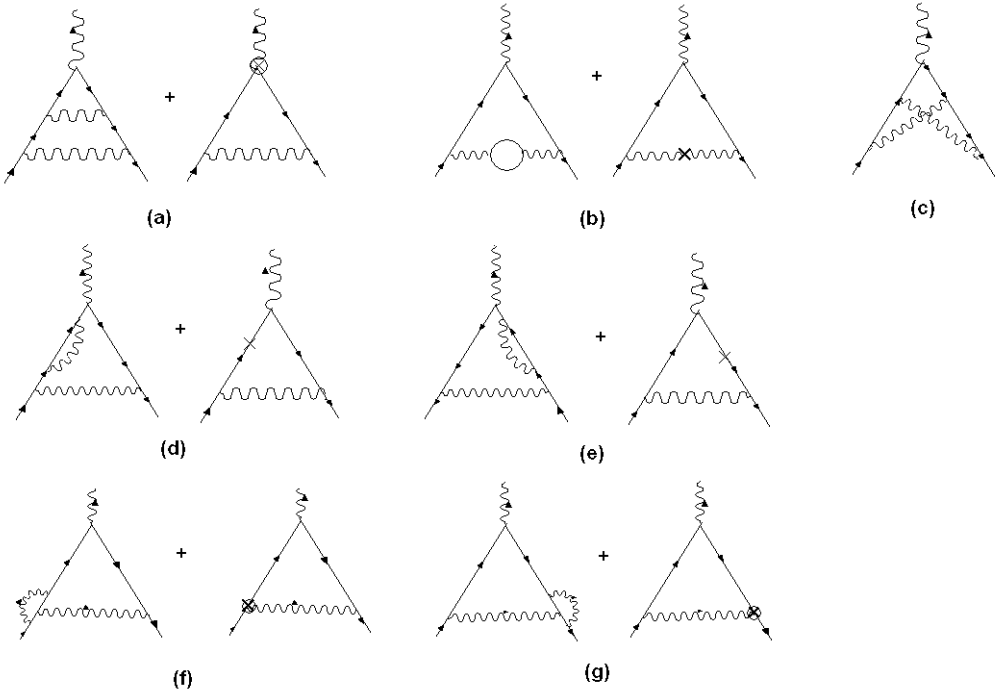}}
  \caption{contributions for the vertex function at spinorial QED}
  \label{fig:5}
\end{figure*}

Now, we verify the Ward Identity which
relates the fermion-photon vertex and the fermion self-energy corrections. In every
order of  perturbation theory, they must obey
\begin{equation}
\label{eq: WI} q^{\mu}\Gamma^{(n)}_{\mu} = \Sigma^{(n)}(p) -
\Sigma^{(n)}(p')\quad \mbox{,}
\end{equation}
where $q=p-p'$ is the momentum of the external photon.
We start by evaluating the two-loop fermion self-energy, for which
picture \ref{fig:4} represents all non-null contributions.

The amplitude of fig.\ref{fig:4}a has been already evaluated in section II and is given by
\be
i\Sigma^{(2)}_a(p)= -2ie^4b \gamma^\mu \left( I_\mu^{(2)} -p_\mu I\right).
\ee
We turn ourselves to the next two-loop self-energy contribution, which is represented
by fig.\ref{fig:4}b, where we can notice that the subdivergence is due to the presence
of the one-loop vacuum polarization tensor.
The expression for this diagram is
\be
\imath \Sigma^{(2)}_b(p) = \imath
 e^4\int^{\Lambda}_k\left\{\gamma^\sigma\frac{1}{\pbruto - \kbruto}\gamma^\rho
\tilde \Pi^{(1)}_{\rho \sigma}  \frac{1}{k^4}\right\},
\ee
where $\tilde \Pi^{(1)}_{\rho \sigma}$ is the finite part of the one-loop vacuum polarization tensor.
It is simple to show that
\begin{eqnarray}
\label{eq: sigmatres}
&&\imath \Sigma^{(2)}_b(p) \nonumber \\
&&=\frac{4}{3}\imath e^4 b\int^{\Lambda}_k \frac{\kbruto(\pbruto -
\kbruto)\kbruto}{k^4(p - k)^2}
\left[\ln\left(-\frac{k^2}{\lambda^2}\right) + \frac{5}{3}\right] \nonumber\\
&&+\frac{4}{3}\imath e^4 b \gamma^\mu \left\{ 2 p_\mu I^{(2)}-2 I^{(2)}_{\mu}
+\frac{5}{3}p_\mu I \right\}.
\end{eqnarray}
%For our purpose of verifying the Ward identities, it is not necessary to evaluate
%the unsolved integral in the above expression, although it is not difficult.

The overlapped diagram in fig. \ref{fig:4}c, with counterterms included
is given by
\be
\imath \Sigma^{(2)}_c=\imath \Sigma^{(2)}_{c1}+2\imath \Sigma^{(2)}_{cCT}
\ee
The contribution of the first term can be easily shown to be
\begin{eqnarray}
&&\imath \Sigma^{(2)}_{c1}(p) = \nonumber\\
&& 4\imath e^4\gamma^{\mu}\left\{-\frac{1}{2}[I_{log}(\lambda^{2}) +
2b]p_{\mu}I(p^2) + bI^{(2)}_{\mu}(p^2)\right\}. \nonumber\\
\end{eqnarray}
The two counterterms give identical contributions
\be
2\imath \Sigma^{(2)}_{cCT}(p)=2ie^4 \pbruto I_{log}(\lambda^2) I,
\ee
and, so
\be
\label{eq: sigmadois}
\imath \Sigma^{(2)}_c(p)= 4\imath e^4
 b\gamma^{\mu}[-p_{\mu}I(p^2) + I^{(2)}_{\mu}(p^2)].
\ee
Summing these three self-energy amplitudes, we finally obtain
\begin{eqnarray}
\label{eq: sigmatotal}
&& \Sigma^{(2)}(p)= \nonumber\\
&& \frac{4}{3} e^4 b\int^{\Lambda}_k \frac{\kbruto(\pbruto -
\kbruto)\kbruto}{k^4(p - k)^2}
\left[\ln\left(-\frac{k^2}{\lambda^2}\right) +
\frac{5}{3}\right]
\nonumber\\
&&+ \frac 23 e^4b \gamma^\mu \left\{ -I_\mu^{(2)}+4p_\mu I^{(2)} +\frac 13 p_\mu I \right\}.
\end{eqnarray}

For the evaluation of the vertex corrections, we would like to call the reader's attention to the fact that,
although Implicit Regularization does not commute with the contraction of the amplitude with the metric
(in this process we would loose control of the surface terms, which possibly are symmetry violating terms),
it commutes with the contraction with an external momentum. For this reason, for the purpose of verifying
the Ward identities, we shall evaluate the vertex corrections contracted
with the momentum of the external photon, $q=p-p'$. The non-null diagrammatic contributions and
the correspondent counterterms to eliminate subdivergences are presented in fig. \ref{fig:5}.

For the diagram of fig.\ref{fig:5}a, we have
\begin{eqnarray}
&&-\imath e \Gamma^{(2)}_{\mu a} = ie^5\int^{\Lambda}_{k,l}
\gamma_\rho\frac{1}{\pbruto' - \kbruto} \gamma_\sigma\frac{1}{\pbruto' - \lbruto}\nonumber\\
&&\times \gamma_\mu\frac{1}{\pbruto - \lbruto}\gamma_\beta\frac{1}{\pbruto -
\kbruto}\gamma_\alpha \frac{g^{\alpha\rho}}{k^2}\frac{g^{\beta\sigma}}{(l- k)^2}\quad \mbox{.}
\end{eqnarray}
It is convenient to isolate the subdiagram, which is the one-loop
vertex correction and contract it with the momentum
of the external photon, $q^\mu$. However, at one-loop order, the
Ward identity relating the vertex function and the fermionic self-energy
is written as $$q^{\mu}\Gamma^{(1)}_{\mu} = \Sigma^{(1)}(p) - \Sigma^{(1)}(p')\quad
 \mbox{,}$$ where $\Sigma^{(1)}(p)$ is the one-loop self-energy of a fermion with
momentum $p$. This allows us to write
\begin{eqnarray}
-\imath e q^{\mu}\Gamma^{(2)}_{\mu a} &=& -\imath e^2
 \int^{\Lambda}_k\gamma^{\alpha}\frac{1}{\pbruto^{'} - \kbruto}(-\imath e)\nonumber\\
                                      & & (-\imath
 e)\left[\Sigma^{(1)}(p - k) - \Sigma^{(1)}(p' - k)\right]\nonumber\\
                                      & & \frac{1}{\pbruto -
 \kbruto}\gamma_{\alpha}\frac{1}{k^2}\mbox{,}
\end{eqnarray}
and, substituting the expression for $\Sigma^{(1)}(p)$, we have
\begin{eqnarray}
\label{eq: vert1}
&&-\imath e q^{\mu}\Gamma^{(2)}_{\mu a} = \nonumber\\
&&2\imath e^5 b\elemint \frac{\pbruto' - \kbruto}{(p' - k)^2k^2}
                                          \left[\ln\left(-\frac{(p
 - k)^2}{\lambda^2}\right) - 2\right]  \nonumber\\
&&-2\imath e^5 b\elemint \frac{\pbruto - \kbruto}{(p - k)^2k^2}
                                          \left[\ln\left(-\frac{(p' -
 k)^2}{\lambda^2}\right) - 2\right].   \nonumber\\
\end{eqnarray}
These integrals will be left unsolved, since they will be canceled by contributions coming from
diagrams of figures \ref{fig:5}d and \ref{fig:5}e, which are presented with their correspondent
counterterms. It is straightforward to show that
\bq
&&-\imath e q^{\mu}\Gamma^{(2)}_{\mu d} = \nonumber\\
&&-2\imath e^5 b\int^{\Lambda}_k\frac{(\pbruto' - \kbruto)}{(p' -
k)^2 k^2}\left[\ln\left(-\frac{(p -k)^2}{\lambda^2}\right) - 2\right]  \nonumber\\
&&+2 \imath e^5 b\int^{\Lambda}_k \frac{(\pbruto - \kbruto)}{(p - k)^2k^2}
\left[\ln\left(-\frac{(p -k)^2}{\lambda^2}\right) - 2\right]\nonumber \\
&&=-2\imath e^5 b\int^{\Lambda}_k\frac{(\pbruto' - \kbruto)}{(p' -
k)^2 k^2}\left[\ln\left(-\frac{(p -k)^2}{\lambda^2}\right) - 2\right]  \nonumber\\
&&+2 \imath e^5 b \gamma^\mu \left\{ I_\mu^{(2)}(p)-p_\mu I(p)\right\}
\eq
and
\bq
&&-\imath e q^{\mu}\Gamma^{(2)}_{\mu e} = \nonumber\\
&&-2\imath e^5 b\int^{\Lambda}_k\frac{(\pbruto' - \kbruto)}{(p' -
 k)^2 k^2}\left[\ln\left(-\frac{(p' - k)^2}{\lambda^2}\right) -
 2\right]\nonumber\\
&&+2\imath e^5 b\int^{\Lambda}_k \frac{(\pbruto - \kbruto)}{(p - k)^2 k^2}
\left[\ln\left(-\frac{(p'- k)^2}{\lambda^2}\right) -
2\right]\nonumber\\
&&= 2\imath e^5 b\int^{\Lambda}_k \frac{(\pbruto - \kbruto)}{(p - k)^2 k^2}
\left[\ln\left(-\frac{(p'- k)^2}{\lambda^2}\right) -2\right] \nonumber \\
&&-2 \imath e^5 b \gamma^\mu \left\{ I_\mu^{(2)}(p')-p'_\mu I(p')\right\}.
\eq

So, in the partial summation,
\bq
&&-\imath e q^{\mu}\left( \Gamma^{(2)}_{\mu a} +\Gamma^{(2)}_{\mu d}+\Gamma^{(2)}_{\mu e}\right)= \nonumber \\
&&2 \imath e^5 b \gamma^\mu \left\{I_\mu^{(2)}(p)-p_\mu I(p)
-\left(I_\mu^{(2)}(p')-p'_\mu I(p')\right)\right\}, \nonumber \\
\eq
occurs the cancelation of the unsolved integrals corresponding to these diagrams.

The next two-loop vertex correction is represented by the diagram of fig.\ref{fig:5}b,
where the subdiagram is a correction for the one-loop photon
propagator. This amplitude is given by
\begin{eqnarray}
&&-\imath e \Gamma^{(2)}_{\mu b} =\nonumber\\
&&(-\imath e)^3 \int^{\Lambda}_k \gamma_{\rho}\frac{\imath}{\pbruto^{'}
 - \kbruto}\gamma_{\mu}
                                   \frac{\imath}{\pbruto -
 \kbruto}\gamma_{\sigma}\left(\frac{-\imath g^{\alpha \sigma}}{k^2}\right)\nonumber\\
&& \left[(-\imath e)^2 (-1)\int^{\Lambda}_l
 Tr\left\{\gamma_{\alpha}\frac{\imath}{\lbruto}\gamma_{\beta}
                                   \frac{\imath}{\lbruto -
 \kbruto}\right\}\right]\left(\frac{-\imath g^{\beta\rho}}{k^2}\right)\mbox{,}\nonumber\\
\end{eqnarray}
where, in the brackets, we have the expression for the subdiagram.
Applying the forest formula and using the expression for the
one-loop correction to the photon propagator we have
\begin{eqnarray}
&&-\imath e \Gamma^{(2)}_{\mu b} = \nonumber\\
&&\imath e^3 \int^{\Lambda}_k \gamma^{\beta}\frac{1}{\pbruto' -
 \kbruto}\gamma_{\mu}\frac{1}{\pbruto -
 \kbruto}\gamma^{\alpha}\frac{1}{k^2}\nonumber\\
&& \left\{-\frac{4}{3}e^2 b [k_{\alpha}k_{\beta} - k^2g_{\alpha\beta}]
 \left[\ln\left(\frac{-k^2}{\lambda^2}\right) + \frac{5}{3}\right]\right\}\frac{1}{k^2}\mbox{.}\nonumber\\
\end{eqnarray}
Contracting the amplitude with the external photon momentum $q^\mu$
and using the identity $\qbruto = (\pbruto - \kbruto) - (\pbruto' -\kbruto)\quad$,
we have, in terms of typical one and two-loop basic integrals,
\begin{eqnarray}
&&-\imath e q^{\mu}\Gamma^{(2)}_{\mu b} =\nonumber\\
&&-\frac 43 \imath b e^5\int^{\Lambda}_k \frac{\kbruto(\pbruto' -
 \kbruto)\kbruto}{k^4(p' -
 k)^2}\left[\ln\left(-\frac{k^2}{\lambda^2}\right) + \frac{5}{3}\right]\nonumber\\
&&+\frac 43 \imath b e^5\int^{\Lambda}_k \frac{\kbruto(\pbruto -
\kbruto)\kbruto}{k^4(p - k)^2}
\left[\ln\left(-\frac{k^2}{\lambda^2}\right) + \frac{5}{3}\right]\nonumber\\
&&+\frac{4}{3}\imath e^4 b \gamma^\mu \left\{ 2 p_\mu I^{(2)}(p)-2 I^{(2)}_{\mu}(p)
+\frac{5}{3}p_\mu I(p) \right\}    \nonumber \\
&&-\frac{4}{3}\imath e^4 b \gamma^\mu \left\{ 2 p'_\mu I^{(2)}(p')-2 I^{(2)}_{\mu}(p')
+\frac{5}{3}p'_\mu I(p') \right\}. \nonumber \\
\end{eqnarray}

The next amplitude is represented by the fig.\ref{fig:5}c and,
following similar steps, the final result for this amplitude can be
written as
\begin{eqnarray}
&&-\imath e q^{\mu} \Gamma^{(2)}_{\mu c} = \nonumber\\
&&-8\imath e^5 \int^{\Lambda}_{k,l} \frac{(\lbruto - \kbruto) l\cdot(p -
 k)}{k^2l^2(k - l)^2(p - k)^2(p'-l)^2}\nonumber\\
&& + 8\imath e^5 \int^{\Lambda}_{k,l} \frac{(\lbruto - \kbruto)
 l\cdot(p' - k)}{k^2l^2(k - l)^2(p' - k)^2(p - l)^2}. \nonumber\\
\end{eqnarray}
To obtain the result above, it was necessary to make shifts and to use the fact that
the integrals are odd in the exchange of both sign of $p$ and $p'$. We do not need to calculate
these integrals, as, as we will see, they will be canceled by some of the next contributions.

Finally, the two last two-loop vertex corrections contributions are
given by the figures \ref{fig:5}f and \ref{fig:5}g.
Calculating the amplitude for the first diagram, we have
\begin{eqnarray}
&&-\imath e q^{\mu} \Gamma^{(2)}_{\mu f} =\nonumber\\
&&-8\imath e^5 \int^{\Lambda}_{k,l} \frac{(\lbruto - \kbruto)l\cdot (p'- k)}{k^2l^2(k-l)^2(p'-k)^2(p-l)^2}\nonumber\\
&&+8\imath e^5 \int^{\Lambda}_{k,l}\frac{(\lbruto -\kbruto)l\cdot (p - k)}{k^2l^2(k-l)^2(p-k)^2(p-l)^2}\nonumber \\
&& -\imath e q^{\mu} \Gamma^{(2)}_{\mu fCT}    \nonumber \\
&&=-8\imath e^5 \int^{\Lambda}_{k,l} \frac{(\lbruto - \kbruto)l\cdot (p'- k)}{k^2l^2(k-l)^2(p'-k)^2(p-l)^2}\nonumber\\
&&-4ie^5b \gamma^\mu \left\{I_\mu^{(2)}(p)-p_\mu I(p)\right\},
\end{eqnarray}
where for the sake of subtracting the subdivergences we have discarded the divergent part of the
subdiagram.
For the last graph, we obtain
\begin{eqnarray}
&&-\imath e q^{\mu} \Gamma^{(2)}_{\mu g} =\nonumber\\
&&8\imath e^5 \int^{\Lambda}_{k,l} \frac{(\lbruto - \kbruto)l\cdot (p- k)}{k^2l^2(k-l)^2(p-k)^2(p'-l)^2}\nonumber\\
&&-8\imath e^5 \int^{\Lambda}_{k,l}\frac{(\lbruto -\kbruto)l\cdot (p' - k)}{k^2l^2(k-l)^2(p'-k)^2(p'-l)^2}\nonumber \\
&&-\imath e q^{\mu} \Gamma^{(2)}_{\mu gCT}   \nonumber \\
&&=8\imath e^5 \int^{\Lambda}_{k,l} \frac{(\lbruto - \kbruto)l\cdot (p- k)}{k^2l^2(k-l)^2(p-k)^2(p'-l)^2}\nonumber\\
&&+4ie^5b \gamma^\mu \left\{I_\mu^{(2)}(p')-p'_\mu I(p')\right\},
\end{eqnarray}
Summing the results for the graphs of figures \ref{fig:5}c, \ref{fig:5}f and \ref{fig:5}g, we get
\begin{eqnarray}
&&-\imath e q^{\mu} (\Gamma^{(2)}_{\mu c} + \Gamma^{(2)}_{\mu f} +
\Gamma^{(2)}_{\mu g}) =\nonumber\\
&&-4ie^5b \gamma^\mu \left\{I_\mu^{(2)}(p)-p_\mu I(p) \right. \nonumber \\
&& \left.- \left(I_\mu^{(2)}(p')-p'_\mu I(p')\right)\right\}
\end{eqnarray}
It is now an easy task to check that
\be
q^\mu \Gamma^{(2)}_\mu= \Sigma^{(2)}(\pbruto)-\Sigma^{(2)}(\pbruto')
\ee

The calculation we have performed above with the contraction of the photon momentum with
the photon-fermion vertex follows the traditional diagrammatic proof of gauge invariance and, thus
it is not surprising that it would work . Nevertheless, it is important to pay attention to this
feature of Constrained Implicit Regularization: the method was developed to contain all the
features which are necessary in order to preserve gauge invariance and dimension sensitive
symmetries. The possibility of making shifts is implemented by taking care of all the surface terms.
The maintenance of the specific dimension of the theory and the fact that the integrand is not modified
guaranties the preservation of the vector algebra. So, the {\it implicit regularization} can even be
a simple cutoff, since its bad features are under control.

In the next section, we perform the calculation of the two loop QED $\beta$-function.
It is a nice test for the definition of counterterms in terms of our basic divergences.
As we have seen, the CIR procedure has preserved the two-loop QED Ward Identities,
which means the correct relations between QED renormalization constants.

\section{Spinorial QED two-loop $\beta$-function}

As a final checking, we are going to present the two-loop spinorial QED $\beta$-function.
The relationship between the bare and renormalized charge is given by
\be
e_B = e \frac{Z_1}{Z_2Z_3^{1/2}}.
\label{renorm}
\ee
As a consequence of the Ward Identities, $Z_1=Z_2$, and it is obtained
through the electromagnetic field counterterm, $Z_3$.
Until two loop order, this counterterm is written as
\begin{eqnarray}
&&Z_3 = 1 + \frac{4}{3}\imath e^2 I_{log}(\lambda^2) - \frac{8}{3}e^4\left\{\frac{3}{2}I^{2}_{log}(\lambda^2) \right. \nonumber\\
&& \left.  -3b I^{(2)}_{log}(\lambda^2) + \frac{31}{6}b I_{log}(\lambda^2)\right\}\quad \mbox{.}
\end{eqnarray}
We write eq. (\ref{renorm}) in terms of the fine structure constant:
\be
\label{eq: barecoup}
\alpha_B = Z^{-1}_3\alpha \quad \mbox{.}
\ee
We are going to use the $\beta$-function definition as
\be
\beta= \frac{\partial (\ln \alpha)}{\partial( \ln \lambda)}=\frac{2 \lambda^2}{\alpha}
\frac {\partial \alpha}{\partial \lambda^2}.
\ee
The expression for $Z_3$ can be written in terms of the coupling constant as
\begin{eqnarray}
&&Z^{-1}_3 = 1 - \frac{16\pi}{3}\imath \alpha I_{log}(\lambda^2) + \frac{8}{3}(4\pi)^2\alpha^2\nonumber\\
&& \times \left(\frac{5}{6}I^2_{log}(\lambda^2) - 3b I^{(2)}_{log}(\lambda^2) + \frac{9}{2}b I_{log}(\lambda^2)\right)+\cdots \nonumber\\
&&= 1 - c_1\alpha + c_2\alpha^2+ \ldots \quad \mbox{.}
\end{eqnarray}
The bare coupling is independent of the scale energy. Therefore, after performing the derivation of
$\alpha_B$ with respect to $\lambda^2$, the scale energy,
we obtain
\be
\frac{\partial \alpha}{\partial \lambda^2}Z^{-1}_3 + \alpha\frac{\partial Z^{-1}_{3}}{\partial\lambda^2}=0
\ee
The first term is given by
\begin{eqnarray}
\label{eq: eqdif}
&&\frac{\partial Z^{-1}_{3}}{\partial\lambda^2}
= -\frac{\partial \alpha}{\partial \lambda^2}c_1 - \frac{\partial c_1}{\partial \lambda^2}\alpha +
2\alpha \frac{\partial \alpha}{\partial \lambda^2}c_2 \nonumber\\
&& + \alpha^2\frac{\partial c_2}{\partial \lambda^2} + \cdots
\end{eqnarray}
After some algebra, we obtain
\begin{eqnarray}
&&\beta = \frac{1}{[1 - 2\alpha c_1 + 3\alpha^2 c_2^2+\cdots]}\left(2\lambda^2\alpha\frac{\partial c_1}{\partial \lambda^2} \right. \nonumber\\
&& \left. -2\lambda^2 \alpha^2 \frac{\partial c_2}{\partial \lambda^2}+\cdots\right).
\end{eqnarray}
We can write
\be
\label{eq: betaeq}
\beta = 2\lambda^2\alpha\frac{\partial c_1}{\partial \lambda^2} -
2\lambda^2\alpha^2\left[\frac{\partial c_2}{\partial \lambda^2} - 2c_1\frac{\partial c_1}{\partial \lambda^2}\right]+\cdots .
\ee
The derivatives of $c_1$ and $c_2$ are given by
\be
\frac{\partial c_1}{\partial \lambda^2} = \frac{\partial}{\partial \lambda^2}\left[\frac{16 \pi i I_{log}(\lambda^2)}{3}\right]
= \frac{1}{3 \pi \lambda^2}
\ee
and
\begin{equation}
\frac{\partial c_2}{\partial\lambda^2} = \frac{8}{3}16\pi^2\left\{\frac{\imath}
{12 \pi^2 \lambda^2}I_{log}(\lambda^2) + \frac{3}{2\pi^4\lambda^2(16)^2}\right\},
\end{equation}
which leads us to
\begin{eqnarray}
&&\frac{\partial c_2}{\partial\lambda^2} - 2c_1\frac{\partial c_1}{\partial \lambda^2} = \nonumber\\
&&\frac{8}{3}16\pi^2\left\{\frac{\imath}{12 \pi^2 \lambda^2}I_{log}(\lambda^2)
+ \frac{3}{2\pi^4\lambda^2(16)^2}\right\} - \nonumber\\
&&\frac{32\imath}{9\lambda^2}I_{log}(\lambda^2) = \frac{1}{4\pi\lambda^2}\quad \mbox{.}
\end{eqnarray}
Finally, we can write the result for the $\beta$ function obtained up to two loop order:
\be
\beta = \frac{2}{3}\frac{\alpha}{\pi} + \frac{1}{2}\left(\frac{\alpha}{\pi}\right)^2\quad \mbox{,}
\ee
which is in agreement with the known result, since the $\beta$ function
is independent of the renormalization scheme up to two loop order.

\section{Conclusion}

In this paper we extended the procedure of Constrained Implicit Regularization to be
applied in abelian gauge theories beyond one-loop order. For the sake of simplicity,
we have considered the massless case. The massive case does not present any new features,
apart from the fact that the calculation is a little more involved. The whole process is based on the elimination of symmetry violating
terms, which are momentum space surface terms. This is equivalent to allow shifts in the momentum
of integration. This is one of the principal requirements in the diagrammatic proof of gauge
invariance and is the reason why Dimensional Regularization preserves this symmetry.

In the case of anomalies, the elimination of the surface terms by means of symmetry restoring
counterterms does not work. The reason is that the counterterm necessary to restore, for instance,
the vector Ward identity would cause the violation of the axial Ward identity and vice-versa. Actually,
this occurs because, diagrammatically, the anomaly manifests itself as a violation of momentum
routing invariance. For methods like Dimensional Regularization, which eliminate surface terms
automatically, the ambiguity is manifested by the form the $\gamma_5$ matrix is inserted
into the trace.

For the non-abelian case, CIR was successfully applied in the
renormalization on QCD to one-loop order \cite{IR10}. We argue here that it is also adequate for
higher order calculations. Our argument is based on the fact that the 't Hooft identity \cite{thooft}
requires a regularization method which allows shifts in the integration momenta. This is well discussed in the references
\cite{taylor} and \cite{cvita}. Implicit Regularization contains all the ingredients to preserve the generalized
Ward-Slavnov-Taylor identities and we expect it would succeed in higher order calculations also in
the non-abelian case as well. That is because after group theoretical factors have been evaluated,
we are left with the same basic integrals, just as in the abelian case.

Actually, the main utility of Implicit Regularization is its application in dimension specific
theories such as supersymmetric theories.
As discussed in the introduction, from a more phenomenological perspective it is expected that
the Minimal Supersymmetric Standard Model (MSSM) can be probed at
LHC with an accuracy at the percent level in measurements of  electroweak precision observables \cite{EWPO}. The theoretical evaluation of such
observables must be performed at least to two loop order both in the SM and MSSM so that the evidence of new physics becomes
unraveled. Clearly, an invariant regularization and renormalization procedure  is important, both to eliminate
inconsistencies and to reduce the number of adjustable parameters in the calculation.

\section{Appendix}
\label{ap: A}
In this appendix we display the results of the basic integrals up
to two loop order and illustrate the procedure with the evaluation of one of them, including the finite part, which
is similar for the others integrals. We begin by the explicit calculation of the two-loop
integral,
\bq
I^{(2)} = \int_k^\Lambda \frac{1}{k^2(k - p)^2}\ln\left(-\frac{k^2}{\lambda^2}\right),
\eq
in which we will insert a fictitious mass $m$ that will be set to zero in the end of the calculation. The use of
the identity (\ref{ident}) gives us
\bq
&&I^{(2)}= \int_k^\Lambda \frac{1}{(k^2-m^2)^2}\ln\left(-\frac{k^2-m^2}{\lambda^2}\right)  \nonumber \\
&&- \int_k \frac{p^2-2 p \cdot k}{(k^2-m^2)^2[(p-k)^2-m^2]}\ln\left(-\frac{k^2-m^2}{\lambda^2}\right) \nonumber \\
&&= I_{log}^{(2)}(m^2) - \tilde I^{(2)}.
\eq
It is clear from the equation above the reason for the introduction of the fictitious mass $m$. Although the
integral by itself is infrared finite, when the separation by means of the relation (\ref{ident}) is performed,
we are left with two infrared divergent parts. The scale relation,
\bq
&&I^{(2)}_{log}(m^2)=I^{(2)}_{log}(\lambda^2) \nonumber \\
&&-b\left\{ \frac 12 \ln^2{\left(\frac{m^2}{\lambda^2}\right)}
+ \ln{\left(\frac{m^2}{\lambda^2}\right)}\right\}, \nonumber
\eq
will make the connection between these parts in order to make the limit $m^2 \to 0$ possible. We show
a simple trick to calculate the finite part. First, we make use of the standard limit
\be
\ln a = \lim_{\epsilon \rightarrow 0}\frac{a^{\epsilon }- 1}{\epsilon}
\ee
to write
\be
\tilde I^{(2)}=\lim_{\epsilon \rightarrow 0} \frac {1}{\epsilon}
\sum_{l=0}^1 A_l,
\ee
with
\be
A_l= \frac{(-1)^{1-l}}{(-\lambda^2)^{l \epsilon}} \int_k \frac{p^2-2 p \cdot k}{(k^2-m^2)^{2-l \epsilon}[(p-k)^2-m^2]}.
\ee
The $A_l$ terms can be calculated by means of Feynman parametrization. As it is easy to note, only the coefficient
of $\epsilon$ in the series for $\epsilon \to 0$ of $A_1$ will contribute. So, we get, for small $m$,
\bq
&&\tilde I^{(2)}= -b\left\{ \frac 12 \ln^2{\left(\frac{m^2}{\lambda^2}\right)}
+ \ln{\left(\frac{m^2}{\lambda^2}\right)}\right. \nonumber \\
&& \left.-\frac{1}{2}\ln^2\left(-\frac{p^2}{\lambda^2}\right) + \ln\left(-\frac{p^2}{\lambda^2}\right)\right\}.
\eq
When the divergent and finite parts are added together and the scale relation is used, the dependence on the
fictitious mass $m$ disappears and we get
\be
I^{(2)} = I^{(2)}_{log}(\lambda^2) + b\left\{-\frac{1}{2}\ln^2\left(-\frac{p^2}{\lambda^2}\right) + \ln\left(-\frac{p^2}{\lambda^2}\right)\right\}
\ee

The results of the other integrals, as defined in the text, are the following:

\begin{equation}
\label{eq: I1}
I = \Ilog - b\lnp + 2b \mbox{;}
\end{equation}
\begin{equation}
\label{eq: Imu1}
I_{\mu} = \frac{p_{\mu}}{2}\left\{\Ilog - b\lnp + 2b\right\} \mbox{;}
\end{equation}
\begin{eqnarray}
\label{eq: Imunu1}
&&I_{\mu\nu}= \frac{1}{3}\duplop\left\{\Ilog - b\lnp + \frac{11}{6}b\right\}  \nonumber\\
&& +\frac{1}{3}p^2g_{\mu\nu}\left\{-\frac{1}{4}\Ilog + \frac{b}{4}\lnp  - \frac{b}{3}\right\};\nonumber \\
&&
\end{eqnarray}

\begin{eqnarray}
\label{eq: Imu2}
&& I^{(2)}_{\mu} = \frac{p_{\nu}}{2}\left\{ I^{(2)}_{log}(\lambda^2) + \frac{1}{2}I_{log}(\lambda^2) \right. \nonumber\\
&&\left. -\frac{b}{2}\left[\ln^2\left(-\frac{p^2}{\lambda^2}\right) - \ln\left(-\frac{p^2}{\lambda^2}\right) - 3\right]
\right\};
\end{eqnarray}

\begin{eqnarray}
\label{eq: Imunu2}
&& I^{(2)}_{\mu\nu} =\frac{1}{3}p_{\mu}p_{\nu} \left\{ I^{(2)}_{log}(\lambda^2) + \frac{5}{6}I_{log}(\lambda^2)\right. \nonumber\\
&& \left. -b\left[\frac{1}{2}\ln^2\left(-\frac{p^2}{\lambda^2}\right)-\frac{1}{3}\ln\left(-\frac{p^2}{\lambda^2}\right)
- \frac{9}{4}\right]\right\}\nonumber\\
&& +\frac{p^2g_{\mu\nu}}{12} \left\{ -I^{(2)}_{log}(\lambda^2) + \frac{1}{6}I_{log}(\lambda^2) + \right. \nonumber\\
&& \left. \left[\frac{1}{2}\ln^2\left(\frac{p^2}{-\lambda^2}\right) - \frac{11}{6}\ln\left(-\frac{p^2}{\lambda^2}\right)
+ \frac{11}{6}\right] \right\};
\end{eqnarray}

\bq
&&J^{(2)}=\frac{p^2}{2}\left[I_{log}(\lambda^2) - b \ln\left(-\frac{p^2}{\lambda^2}\right)\right. \nonumber\\
&& \left. +3b \right]
\eq

Another class of integrals we have to deal to two loop order are the {\it overlapped} ones. In this case, there is a symmetry between the
internal momentum $k$ and $l$ in the denominator, with three factors involving each one. This type of integrals commonly
appears in the solution of overlapped diagrams. So it is interesting to maintain all the subdivergences to be eliminated at
the end, when the counterterms are added. We give some details of the solution of one of them:

\bq
&&I_{\mu \nu}^{{\cal O}1}=\int_{k,l}^\Lambda \frac{k_\mu k_\nu}{k^2l^2(k-l)^2(p-l)^2(p-k)^2}  \nonumber \\
&&=\int_{k,l}^\Lambda \frac{k_\mu k_\nu}{k^4l^2(k-l)^2(p-l)^2}  \nonumber \\
&&-\int_{k,l} \frac{(p^2-2p \cdot k)k_\mu k_\nu}{k^4l^2(k-l)^2(p-l)^2(p-k)^2} \nonumber \\
&&=A_{\mu \nu}-B_{\mu \nu}
\eq
The second integral is finite. The first one is easily solved:
\bq
&&A_{\mu \nu}=\int_l^\Lambda \frac{1}{l^2(p-l)^2)}\int_k^\Lambda \frac{k_ \mu k_\nu}{k^4(k-l)^2} \nonumber \\
&&=\int_l^\Lambda \frac{1}{l^2(p-l)^2)} \left\{ \frac{g_{\mu \nu}}{4}\left[ I_{log}(\lambda^2)
-b \ln\left(-\frac{l^2}{\lambda^2}\right) \right. \right. \nonumber \\
&& \left. \left.+2b\right] + \frac{b}{2}\frac{l_\mu l_\nu}{l^2}\right\}  \nonumber \\
&& = \frac{g_{\mu \nu}}{4} \left(I_{log}(\lambda^2)+\frac{5}{2}b\right) I -b \frac{g_{\mu \nu}}{4} I^{(2)} \nonumber \\
&&+ \frac{b^2}{4} \frac{p_ \mu p_\nu}{p^2}
\eq
For the finite part, we adopt the following procedure. We write
\be
B_{\mu \nu}=C_1 p_\mu p_\nu + C_2p^2 g_{\mu \nu}
\ee
and by contracting the integral $B_{\mu \nu}$ with $p_\mu p_\nu$ and with $g_{\mu \nu}$, we solve a system of two equations
to obtain $C_1$ and $C_2$. This procedure is not allowed for divergent integrals in the context of Implicit Regularization,
since the contraction with $g_{\mu \nu}$ would make us loose control of the surface terms that must be eliminated.
The final result for $I_{\mu \nu}^{{\cal O}1}$ yields
\bq
&& I_{\mu \nu}^{{\cal O}1} =  \frac{g_{\mu \nu}}{4} \left\{ I_{log}^2(\lambda^2)- b I_{log}^{(2)}(\lambda^2)
+\frac 92 b I_{log}(\lambda^2)  \right.  \nonumber \\
&& \left. -  b \ln\left(-\frac{p^2}{\lambda^2}\right) I_{log}(\lambda^2)
+ \frac {b^2}{2} \ln^2\left(-\frac{p^2}{\lambda^2}\right)
\right. \nonumber \\
&& \left.- \frac 72 b^2 \ln\left(-\frac{p^2}{\lambda^2}\right) + \frac{31}{6} b^2 - \frac{p^2}{3} I^{{\cal O}} \right\}
\nonumber \\
&&+ \frac{p_\mu p_\nu}{p^2} \left\{\frac{p^2}{3} I^{{\cal O}} + \frac {b^2}{12} \right\}.
\eq
For the other divergent integral of this type we needed, we have:
\begin{eqnarray}
\label{eq: ImunuOab}
&&I^{{\cal O}2}_{\mu\nu}= g_{\mu\nu}\left\{ \frac{b}{4}I_{log}(\lambda^2) - \frac{b^2}{4}\lnp - \frac{p^2}{12}I^{\cal O} \right. + \nonumber\\
&& \left. \frac{11}{12}b^2 - b^2\frac{\pi^2}{36}\right\} + \nonumber\\
&&+ \frac{\duplop}{p^2}\left\{\frac{p^2}{3}I^{\cal O} - \frac{1}{6}b^2 + b^2\frac{\pi^2}{36}\right\}.
\end{eqnarray}

\end{document}